\documentclass[jgr, draft]{AGUTeX}
\usepackage[T1]{fontenc}
\usepackage[latin9]{inputenc}
\usepackage{array}
\usepackage{multirow}
\usepackage{amsmath}
\usepackage{amssymb}
\usepackage{graphicx}
\usepackage{esint}

\makeatletter

\providecommand{\tabularnewline}{\\}

\usepackage[small]{caption}

\setkeys{Gin}{draft=false}

\authorrunninghead{ARCHER ET AL.}

\titlerunninghead{STANDING MAGNETOPAUSE SURFACE WAVES}

\authoraddr{M. O. Archer,
Space \& Atmospheric Physics Group, The Blackett Laboratory, Imperial College London, Prince Consort Road, London, SW7 2AZ, UK.
(m.archer10@imperial.ac.uk)}

\authoraddr{F. Plaschke, Space Research Institute, Austrian Academy of Sciences, 8042 Graz, Austria.
(Ferdinand.Plaschke@oeaw.ac.at)}

\makeatother

\begin{document}

\title{What frequencies of standing surface waves can the subsolar magnetopause
support?}

\authors{M. O. Archer, \altaffilmark{1}\\
and F. Plaschke, \altaffilmark{2}}

\altaffiltext{1}{Blackett Laboratory, Imperial College London, London, SW7 2AZ, UK.}

\altaffiltext{2}{Space Research Institute, Austrian Academy of Sciences, 8042 Graz,
Austria.}
\begin{abstract}
It is has been proposed that the subsolar magnetopause may support
its own eigenmode, consisting of propagating surface waves which reflect
at the northern/southern ionospheres forming a standing wave. While
the eigenfrequencies of these so-called Kruskal-Schwartzschild (KS)
modes have been estimated under typical conditions, the potential
distribution of frequencies over the full range of solar wind conditions
is not know. Using models of the magnetosphere and magnetosheath applied
to an entire solar cycle's worth of solar wind data, we perform time-of-flight
calculations yielding a database of KS mode frequencies. Under non-storm
times or northward interplanetary magnetic field (IMF), the most likely
fundamental frequency is calculated to be $0.64_{-0.12}^{+0.03}$~mHz,
consistent with previous estimates and indirect observational evidence
for such standing surface waves of the subsolar magnetopause. However,
the distributions exhibit significant spread (of order $\pm$0.3~mHz)
demonstrating that KS mode frequencies, especially higher harmonics,
should vary considerably depending on the solar wind conditions. The
implications of such large spread on observational statistics are
discussed. The subsolar magnetopause eigenfrequencies are found to
be most dependent on the solar wind speed, southward component of
the IMF and the $Dst$ index, with the latter two being due to the
erosion of the magnetosphere by reconnection and the former an effect
of the expression for the surface wave phase speed. Finally, the possible
occurrence of KS modes is shown to be controlled by the dipole tilt
angle.
\end{abstract}
\begin{article}

\section{Introduction}

Ultra-low frequency (ULF) waves in the Pc5 (2-7~mHz) range play a
significant role in the mass, energy, and momentum transport within
the Earth\textquoteright s magnetosphere e.g. through drift and bounce
resonances with electrons in the outer radiation belt \citep{claudepierre13,mann13}.
Thus it is desirable to be able to predict the locations of magnetospheric
ULF waves and, perhaps more importantly, their frequencies under the
full range of solar wind conditions. The magnetosphere can support
a number of resonantly excited ULF eigenmodes including: Field Line
Resonances (FLRs), standing Alfv\'{e}n waves on local field lines
fixed at their ionospheric ends \citep[e.g.][]{southwood74}; and
cavity or waveguide modes, radially standing fast mode waves trapped
between magnetospheric boundaries or turning points \citep[e.g.][]{kivelson85}.

It has been proposed that the subsolar magnetopause may also support
its own eigenmode which have been referred to as Kruskal-Schwartzschild
(KS) modes, standing magnetopause surface waves or magnetopause surface
eigenmodes \citep{plaschke09a,archer13c}. A theory of these eigenmodes
was developed by \citet{plaschke11}, following on from work by \citet{kruskal54}
and \citet{chen74}, using ideal magnetohydrodynamic (MHD) theory
for incompressible plasmas in a box model magnetosphere. They consist
of a pair of stable propagating surface waves (propagating parallel
and antiparallel to the geomagnetic field) which reflect at the northern
and southern ionospheres \citep{plaschke09a}. The superposition of
these two surface waves results in a standing surface wave of the
magnetopause, as illustrated in Figure~\ref{fig:cartoon}, thereby
quantising the possible resonance frequencies and forming an eigenmode
of the magnetopause. Note that these are unrelated to surface waves
and vortices due to the Kelvin-Helmholtz instability, which typically
occur far down the magnetopause flanks where large velocity shears
are present \citep[e.g. the review of][]{johnson14}.

Standing magnetopause surface waves are thought to only be possible
in the vicinity of the subsolar magnetopause, since the fast magnetosheath
flow acts to convect surface waves tailward \citep[e.g.][]{pu83}.
In order to establish an eigenmode of the magnetopause, the pair of
oppositely propagating surface waves must be allowed to interfere
with one another to form a standing wave. Away from the subsolar magnetopause,
however, it is unlikely that the reflected surface waves would be
able to propagate azimuthally against the opposing magnetosheath flow,
thereby making standing surface waves not possible \citep{plaschke11}.
Similarly, these reasons suggest that KS modes require small azimuthal
wavenumbers.

While the subsolar magnetopause eigenmode shares a number of characteristics
with FLRs, there are key differences. FLRs consist of toroidal or
poloidal mode Alfv\'{e}n waves which are localised to a field line,
hence the eigenfrequencies of FLRs vary with L-shell/latitude \citep[e.g.][]{lee89}.
On the other hand, KS modes consist of magnetopause surface waves,
which are theoretically described in MHD as two evanescent magnetosonic
waves, one on the magnetosheath and the other on the magnetospheric
side, tied together with boundary conditions \citep[e.g.][]{pu83}.
Their magnetospheric signatures thus exponentially decay in amplitude
with distance from the boundary. The eigenfrequencies of KS modes
are characterised by the time it takes a disturbance at the magnetopause
to propagate as a pair of surface waves to the northern/southern ionospheres,
reflect back and then interfere with one another. This is in turn
a function of the physical properties of the magnetosheath and magnetospheric
plasma along the path of the surface waves \citep{chen74}. \citet{plaschke09a}
used typical magnetospheric and magnetosheath conditions at the nose,
yielding an estimated fundamental eigenfrequency of $\sim$0.6~mHz.
Subsequently, \citet{archer13c} used a similar method applied to
130 events showing that the eigenfrequencies (and moreso the surface
wave phase speed) should correlate with the solar wind speed. However,
such estimates of KS mode frequencies to date have not taken into
account the full variability of the magnetospheric system. Furthermore,
the calculations have used highly simplified box models of the magnetosheath-magnetosphere-ionosphere
system whereby the plasma quantities have been assumed spatially constant
and the effects of the magnetosheath flow ignored.

The first indirect observational evidence towards this ULF wave eigenmode
was reported by \citet{plaschke09a} from a statistical set of observed
subsolar magnetopause oscillation frequencies showing greater occurrence
at some discrete frequencies, namely \{1.3, 1.9, 2.7, 3.1, 4.1\}$\pm0.1$~mHz.
From the regularity of the observed frequencies it was inferred that
these could possibly be explained as harmonics (integer multiples)
of a fundamental eigenfrequency $\sim$0.65~mHz hence may be consistent
with the expected KS mode frequencies. Boundary oscillations at these
frequencies were found to occur more often under quasi-radial interplanetary
magnetic field (IMF) \citep{plaschke09}. Magnetosheath jets/dynamic
pressure pulses are known to predominantly originate from the quasi-parallel
bow shock i.e. quasi-radial IMF \citep{archer13b,plaschke13} and
it has been suggested that such localised pressure enhancements might
be a natural driver for standing magnetopause surface waves \citep{plaschke11}.\citet{archer13c}
thus investigated the statistical reponse of the subsolar magnetosphere
to previously identified magnetosheath jets/dynamic pressure pulses
\citep{archer13}. They found that the broadband jets indeed excited
similar discrete frequencies, chiefly in the compressional component
of the magnetic field, at geostationary orbit which they intepreted
as further indirect evidence of KS modes. 

It should be noted that comparable, possibly quasi-steady, discrete
frequencies have been reported in a number of studies \citep{fenrich95,chisham97,francia97,villante01,kokubun13}.
These have been attributed to different modes, depending on the circumstances
of the observations (location, solar wind conditions etc.), such as
cavity/waveguide modes in the flank magnetosphere \citep{samson91,samson92};
waves across the dayside directly driven by solar wind dynamic pressure
oscillations \citep{kepko02,kepko03,viall08}; or pulsed reconnection
at the subsolar magnetopause due to oscillations of the IMF direction
\citep{prikryl98,prikryl99}. Some statistical studies have shown
little evidence for predominant discrete quasi-steady frequencies
though \citep{baker03,rae12}. It is therefore of interest to theoretically
estimate the distribution of possible discrete frequencies due to
the different known modes under the full range of solar wind and magnetospheric
conditions. In this paper we restrict ourselves to just one of these
modes, the eigenmode of the subsolar magnetopause, performing time-of-flight
calculations applied to more representative models of the magnetosphere
and magnetosheath than previous studies. Thus we estimate, for the
first time, the distribution of KS mode eigenfrequencies over an entire
solar cycle, revealing their most likely set of eigenfrequencies and
quantifying how variable these should be in general. We also investigate
what parameters primarily control the possible occurrence and frequency
of these modes, determining the physical explanation for these dependences.

\section{Method}

\subsection{Data \& Models}

\begin{figure}
\begin{centering}
\includegraphics{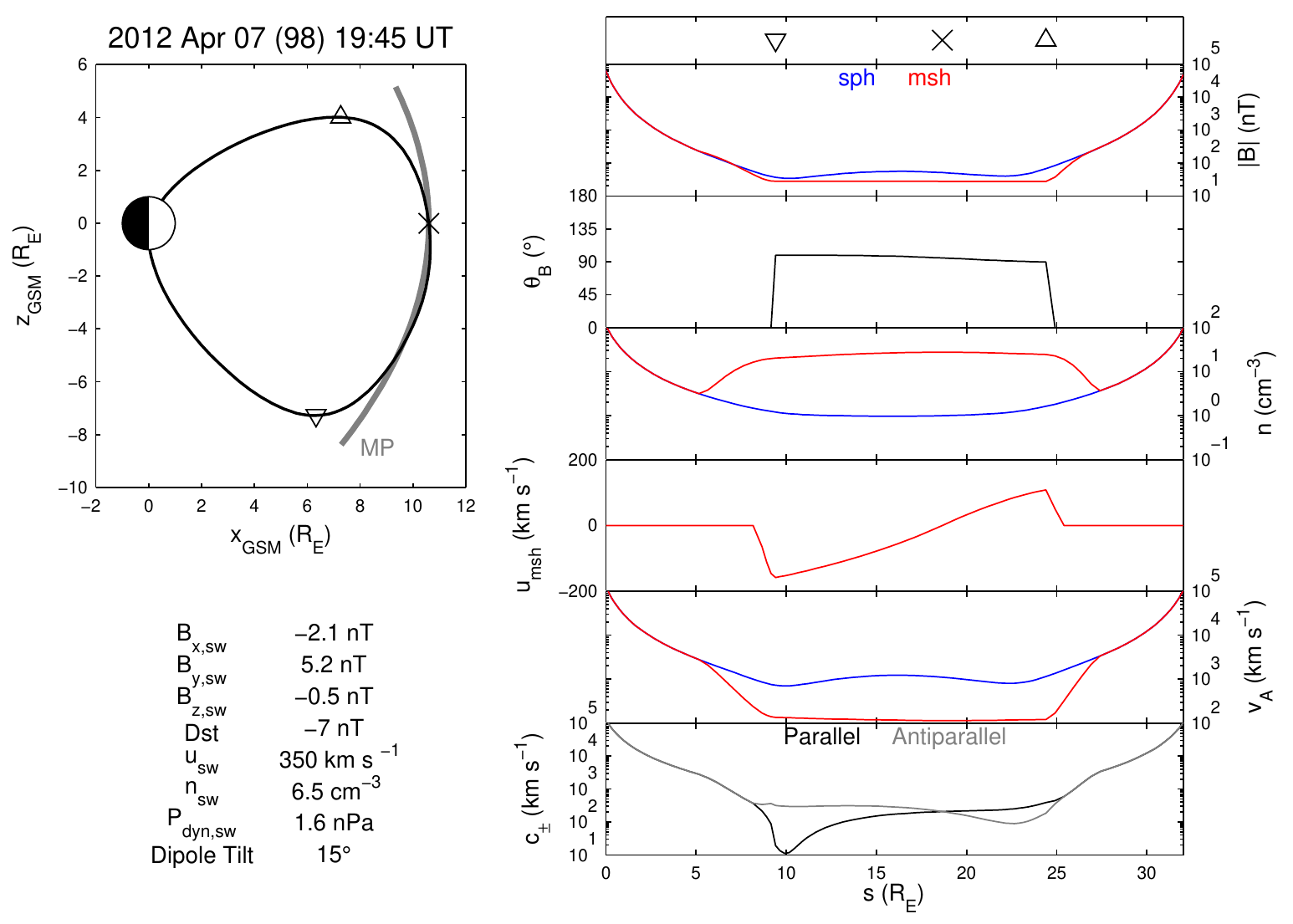}
\par\end{centering}

\protect\caption{Example KS mode frequency calculation.\textbf{ }Left\textbf{:} Subsolar
magnetopause field line (black) from the T96 magnetospheric magnetic
field model in the GSM x-z plane. The subsolar point (cross) and extrema
(triangles) of the field line are also indicated. The paraboloidal
magnetopause used for the KF94 magnetosheath magnetic field model
is also shown (grey). Right: magnetic field strengths; magnetic shear
angle; number densities; magnetosheath flow speeds; Alfv\'{e}n speeds;
and surface wave phase speeds for the parallel (black) and antiparallel
(grey) propagating surface waves. Model quantities are shown as a
function of length along the field line from the southern footpoint
to the northern. Values on the magnetospheric and magnetosheath sides
of the boundary are shown in blue and red respectively. The calculated
fundamental frequency here was $f_{KS}=0.49$~mHz.\label{fig:Example}}
\end{figure}

We use 5~min resolution solar wind data (in GSM coordinates where
appropriate) from the OMNI database between 2001-2013, spanning an
entire solar cycle. OMNI combines observations from numerous spacecraft
to produce an estimate of the solar wind conditions at the bow shock
nose. There were 1,262,690 datapoints during this period. To estimate
the frequencies of standing surface waves that the subsolar magnetopause
may support, we combine a number of physics-based and semi-empirical
models of magnetosheath and magnetospheric properties which are summarised
in Table~\ref{tab:Models}. These models are applied to the subsolar
magnetopause field line (also known as the last closed field line)
from the T96 magnetospheric magnetic field model \citep{t95,t96}.
While the KF94 \citep{KF94} magnetosheath magnetic field model contains
a different explicitly defined boundary to T96, we connect them together
through the solar zenith angle. Furthermore, in the polar cusps the
magnetic field and density are interpolated between the magnetosheath
and magnetospheric values as a function of geocentric distance from
the extrema (triangles in Figure~\ref{fig:Example}) to 60\% of this
distance \citep[c.f.][]{lavraud04}. We also reduce the magnetosheath
flow speed to zero within 10\% of the distance from the extrema.

\subsection{KS Mode Frequency Calculation}

\begin{figure}
\begin{centering}
\includegraphics{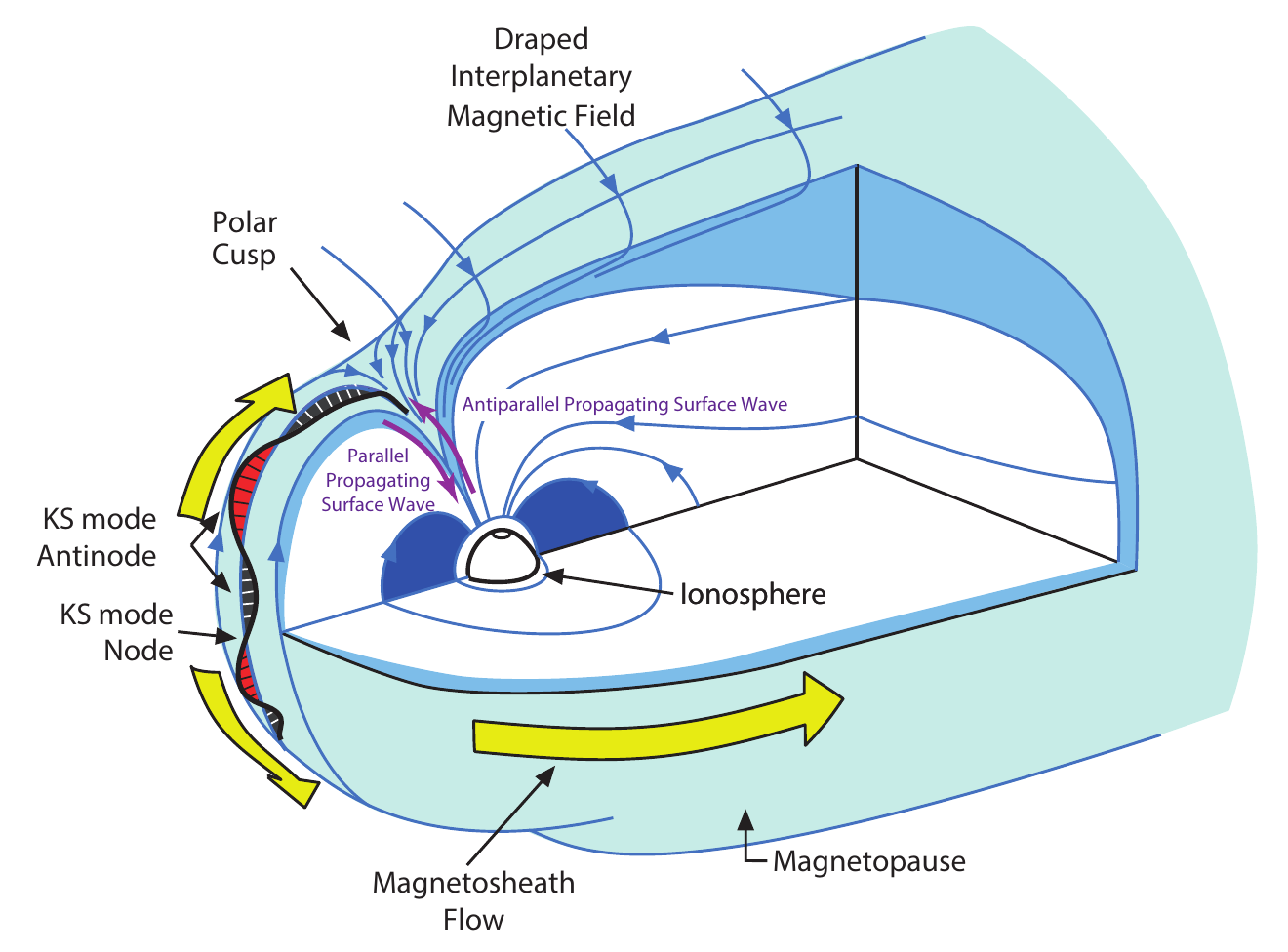}
\par\end{centering}

\protect\caption{Illustration of standing surface waves at the subsolar magnetopause,
known as Kruskal-Schwartzschild (KS) modes. Adapted from \citet{dekeyser05}.\label{fig:cartoon}}
\end{figure}

In this paper we estimate the fundamental frequencies of standing
surface waves at the subsolar magnetopause, assuming that their wavevectors
$\mathbf{k}=k_{\mu}\boldsymbol{\mu}+k_{\nu}\boldsymbol{\nu}+k_{\phi}\boldsymbol{\phi}$
(where $\boldsymbol{\mu}$ is along the T96 geomagnetic field, $\boldsymbol{\nu}$
is normal to the field line pointing outwards and $\boldsymbol{\phi}$
is the usual azimuthal direction) have vanishing azimuthal component
i.e. $k_{\phi}=0$, such that the surface waves do not propagate downtail
(see section \ref{sub:Validity} for more discussion). Using incompressible
MHD, the local dispersion relation for a magnetopause surface wave
is given by \citep[c.f.][]{plaschke11}

\begin{equation}
\left(\rho_{msh}+\rho_{sph}\right)\left(\frac{\omega}{k_{\mu}}\right)^{2}-2\rho_{msh}u_{msh}\left(\frac{\omega}{k_{\mu}}\right)+\rho_{msh}u_{msh}^{2}-\rho_{msh}v_{A,msh}^{2}\cos^{2}\theta_{B}-\rho_{sph}v_{A,sph}^{2}=0\label{eq:dispersion}
\end{equation}
where the subscripts $msh$ and $sph$ correspond to the magnetosheath
and magnetospheric sides of the boundary respectively, $\rho=m_{p}n$
is the mass density assuming a purely proton composition, $u$ is
the velocity, $v_{A}$ is the Alfv\'{e}n speed, $\theta_{B}$ is
the magnetic shear angle between the magnetosheath and magnetospheric
fields, and $\omega$ is the angular frequency of the wave. While
we account for the Doppler effect of the magnetosheath flow, we assume
that velocities inside the magnetosphere are negligible. Since Equation~\ref{eq:dispersion}
is quadratic, there are two analytical solutions to the phase speed
$\omega/k_{\mu}=c_{\pm}$, corresponding to surface waves propagating
parallel ($+$) or antiparallel ($-$) to the geomagnetic field as
illustrated in Figure \ref{fig:cartoon}. Using the T96 model, we
determine positions along the subsolar magnetopause field line and
compute the two phase speeds at each point using the models given
in Table~\ref{tab:Models}. A worked example is given in Figure~\ref{fig:Example}.

Since standing waves must consist of both parallel and anti-parallel
propagating surface waves which reflect at the northern and southern
ionospheres (see Figure~\ref{fig:cartoon}), if the propagation direction
of either of these surface waves is reversed in Earth's rest frame
at some point along the field line (due to the magnetosheath flow)
then a KS mode cannot be supported \citep{plaschke11}. On the other
hand, when this does not occur we arrive at the fundamental standing
surface wave frequency $f_{KS}$ using the time-of-flight technique

\begin{equation}
f_{KS}\equiv\left[\sum_{\pm}\int\frac{ds}{c_{\pm}}\right]^{-1}\label{eq:tof}
\end{equation}
where $ds$ is a differential line element along the field line. We
assess the validity of these calculations given the models and assumptions
used in section \ref{sub:Validity}.

\section{Occurrence}

\begin{figure}
\begin{centering}
\includegraphics{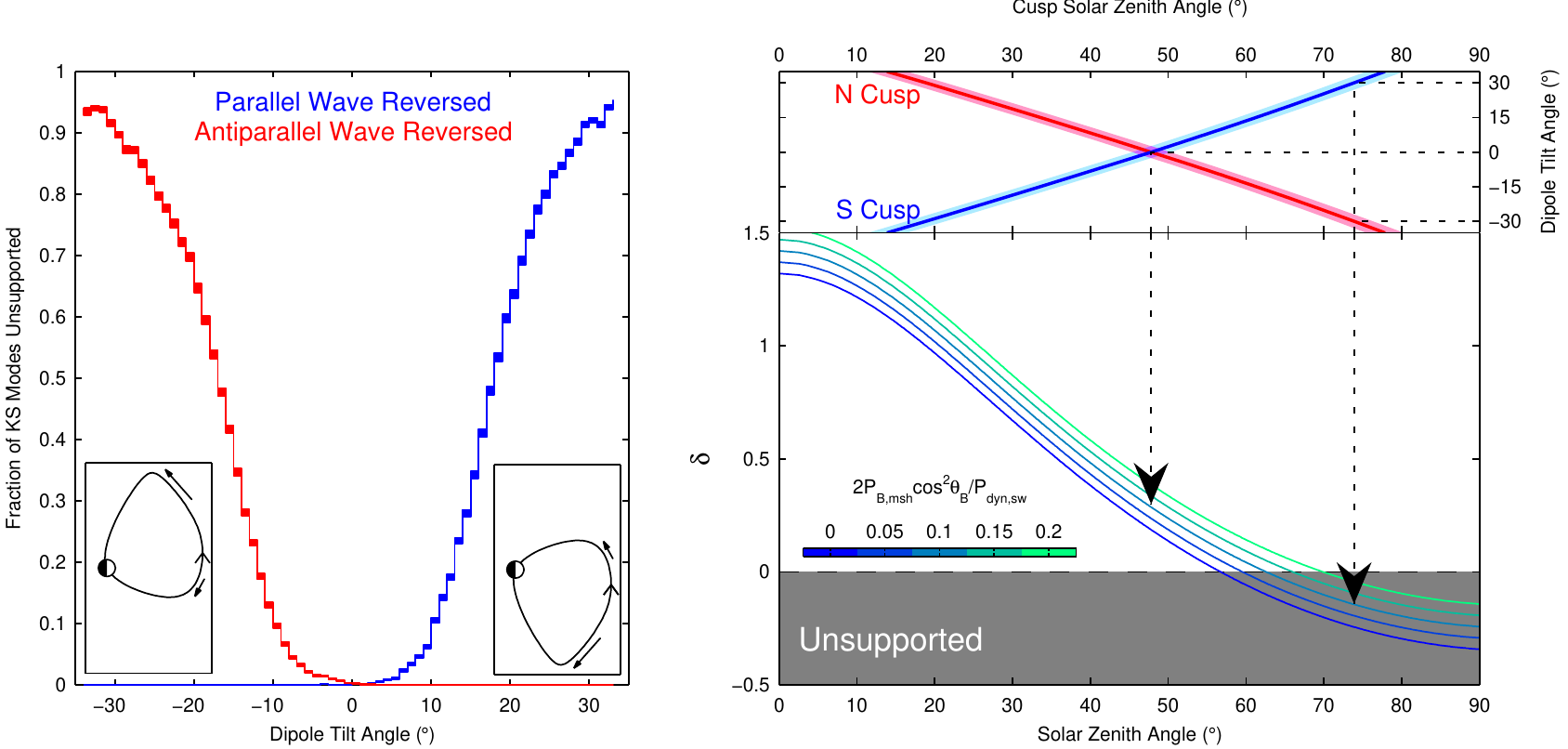}
\par\end{centering}

\protect\caption{Left: Fraction of KS modes unsupported (due to reversal by the magnetosheath
flow) as a function of the dipole tilt angle due to the parallel (blue)
or antiparallel (red) propagating surface waves being reversed by
the magnetosheath flow. The height of the bars indicates the 95\%
confidence intervals. Inset are examples of the T96 subsolar magnetopause
field line in the x-z GSM plane for negative (left inset) and positive
(right inset) tilt, with the magnetosheath flow and geomagnetic field
direction also indicated. Right: The solar zenith angle locations
of the northern (red) and southern (blue) cusps as a function of dipole
tilt angle are shown in the top panel. The lines show results for
a magnetopause standoff distance of $10\,\mathrm{R_{E}}$, with the
shaded regions indicating $\pm1\,\mathrm{R_{E}}$. The bottom panel
shows the variation of the parameter $\delta$ defined in Equation~\ref{eq:reversal3},
whereby $\delta\leq0$ means that KS modes are unsupported, for a
representative range of values of $2P_{B,msh}\cos^{2}\theta_{B}/P_{dyn,sw}$
given by the colour scale.\label{fig:tilt}}
\end{figure}

We investigate the conditions which control the possible occurrence
of standing surface waves at the subsolar magnetopause by plotting
histograms of the fraction of the time that they were unsupported
(due to a reversal of the phase speed by the magnetosheath flow making
the surface wave unable to reach its target ionosphere) as a function
of the model inputs. Overall, KS modes were allowed 61\% of the time
in our model, corresponding to 765,553 computed frequencies. We find
that the main controlling parameter of KS mode occurrence is the dipole
tilt angle, as shown in Figure~\ref{fig:tilt} (left). For small
tilt angles standing surface waves are largely allowed, whereas they
are generally unsupported at large tilt angles. Furthermore, it is
the parallel propagating wave which is reversed (blue) for positive
dipole tilts and the antiparallel wave (red) for negative tilts.

Here we try to understand this dependence on dipole tilt theoretically.
From Equation~\ref{eq:dispersion} it follows that KS modes are unsupported
due to the reversal of either the parallel or antiparallel surface
wave by the magnetosheath flow if at any point along the field line

\begin{subequations}\label{eq:reversal}

\begin{align}
0 & \geq-\rho_{msh}u_{msh}^{2}+\rho_{sph}v_{A,sph}^{2}+\rho_{msh}v_{A,msh}^{2}\cos^{2}\theta_{B}\label{eq:reversal1}\\
\Rightarrow0 & \geq-P_{dyn,msh}+2P_{B,sph}+2P_{B,msh}\cos^{2}\theta_{B}\label{eq:reversal2}\\
\Rightarrow0 & \geq-\frac{P_{dyn,msh}}{P_{dyn,sw}}+\frac{2P_{B,sph}}{P_{dyn,sw}}+\frac{2P_{B,msh}\cos^{2}\theta_{B}}{P_{dyn,sw}}\equiv\delta\label{eq:reversal3}
\end{align}

\end{subequations}where the subscript $sw$ refers to the solar wind,
$P_{dyn}=\rho u^{2}$ is the dynamic pressure, $P_{B}$ is the magnetic
pressure and we introduce the parameter $\delta$ which becomes negative
when one of the waves is reversed. The \citet{spreiter66} models
of the magnetosheath density and flow are proportional to their respective
solar wind conditions, thus the first term in Equation~\ref{eq:reversal3}
is a function of solar zenith angle only. For simplicity, here we
consider the second term using pressure balance \citep[e.g.][]{spreiter66}
applied to the KF94 magnetopause model, which again makes this term
solely dependent on the solar zenith angle. While the third term does
vary with the upstream conditions, it is typically small. We therefore
plot the parameter $\delta$ in Figure~\ref{fig:tilt} (bottom right)
as a function of solar zenith angle for a representative range of
the third term. This shows that $\delta$ decreases with increasing
solar zenith angle, becoming zero at $\sim$60-70$^{\circ}$. Therefore,
if one of the polar cusps are located at a similarly large solar zenith
angle then KS modes will be unsupported due to a reversal of one of
the surface waves.

Figure~\ref{fig:tilt} (top right) shows the solar zenith angles
of the intersection of the northern (red) and southern (blue) cusps
with the KF94 magnetopause model as a function of the dipole tilt
angle, assuming a constant invariant latitude of the polar cusps of
$\pm78{}^{\circ}$ \citep{russell00}. For zero dipole tilt, both
cusps are located at zenith angles of $\sim50{}^{\circ}$ where $\delta$
is positive and thus standing surface waves at the subsolar magnetopause
are supported. For $+30{}^{\circ}$ dipole tilt, the southern cusp
is located at a large solar zenith angle $\sim75{}^{\circ}$ where
$\delta$ has become negative. Here the magnetosheath flow is in opposition
to the parallel propagating wave (as seen in the right inset of Figure~\ref{fig:tilt})
hence this wave is reversed and a KS mode is not possible. Similarly,
for $-30{}^{\circ}$ dipole tilt it is the northern cusp at a large
zenith angle where the magnetosheath flow opposes the anti-parallel
wave (illustrated in the left inset) thereby reversing it. This simple
theoretical treatment is therefore in agreement with the determined
dependence of KS mode occurrence on dipole tilt angle in the full
model.

\section{Frequencies}

\subsection{Results}

\subsubsection{Distributions}

Figure~\ref{fig:distributions} (top left) shows distributions of
the calculated fundamental standing surface wave frequencies of the
subsolar magnetopause. We have separated all the computed frequencies
by geomagnetic activity using a $Dst$ threshold of $-10$~nT, close
to the median value, to distinguish between storm (grey; 393,958 datapoints)
and non-storm (black; 371,595 datapoints) times. In addition we separate
all the frequencies under northward (red; 287,266 datapoints) or southward
(blue; 478,287 datapoints) interplanetary magnetic field (IMF). The
distributions show the correctly normalised probability density functions
(PDFs) i.e. the probability that $f_{KS}$ was between $f$ and $f+df$
is given by $PDF\left(f\right)df$ such that the area under each PDF
is unity.

We find that under non-storm times or northward IMF the most likely
frequency is 0.64~mHz. Given the vast number of samples in our distributions
and the insensitivity of the result with different bin sizes, we have
high statistical confidence that this is the mode, i.e. the most likely
value, of our model calculations. The accuracy of our calculations
are later discussed in section \ref{sub:Validity}. This is consistent
with the $\sim$0.65~mHz fundamental suggested by \citet{plaschke09a}
(indicated by the vertical dashed line). In contrast, during storm
times or under southward IMF the most likely frequency is greater
(more so for the former), hence inconsistent with previous estimates.
Note that the computed fundamental FLR frequencies for the same field
line are much larger, typically by a factor of $\sim$3-6.

The distributions are highly skewed, as evidenced by the medians being
larger than the modes in all cases. Indeed, this must be the case
since negative frequencies are not possible. In fact, the distributions
shown in Figure~\ref{fig:distributions} can be fairly well modelled
as log-normal (not shown). It is clear that all the distributions
show significant spread as evidenced by their interquartile ranges
(IQRs). This spread, which is of the order of $\pm$0.3~mHz ($\sim$50\%
the most likely value) in the cases of non-storm times or northward
IMF and much larger otherwise, is significant. Therefore, while we
quote a most likely frequency of 0.64~mHz, frequencies outside this
range should often occur. For instance in our model distributions
0.64~mHz is only 35\% more likely than either 0.5~mHz or 0.8~mHz
during non-storm times, but $\sim$30 times more likely than 0.1~mHz
or 1.5~mHz.

We also wish to understand the time variability of the subsolar magnetopause
eigenfrequencies. We therefore construct waiting time distributions
for absolute changes in $f_{KS}$ greater than some threshold percentage.
Fitting these to a negative exponential distribution, we find that
20\% changes (used here since it is greater than the presumed accuracy
of the calculations as discussed in section~\ref{sub:Validity})
in the frequency occur after a characteristic timescale of 48~min.
Increasing the frequency change threshold by successive 10\% increments
results in further factor $\sim$1.75 increases in the characteristic
waiting times. We thus conclude that KS mode frequencies should be
relatively stable over the time of wave propagation.

\subsubsection{Dependences}

\begin{figure}
\begin{centering}
\includegraphics{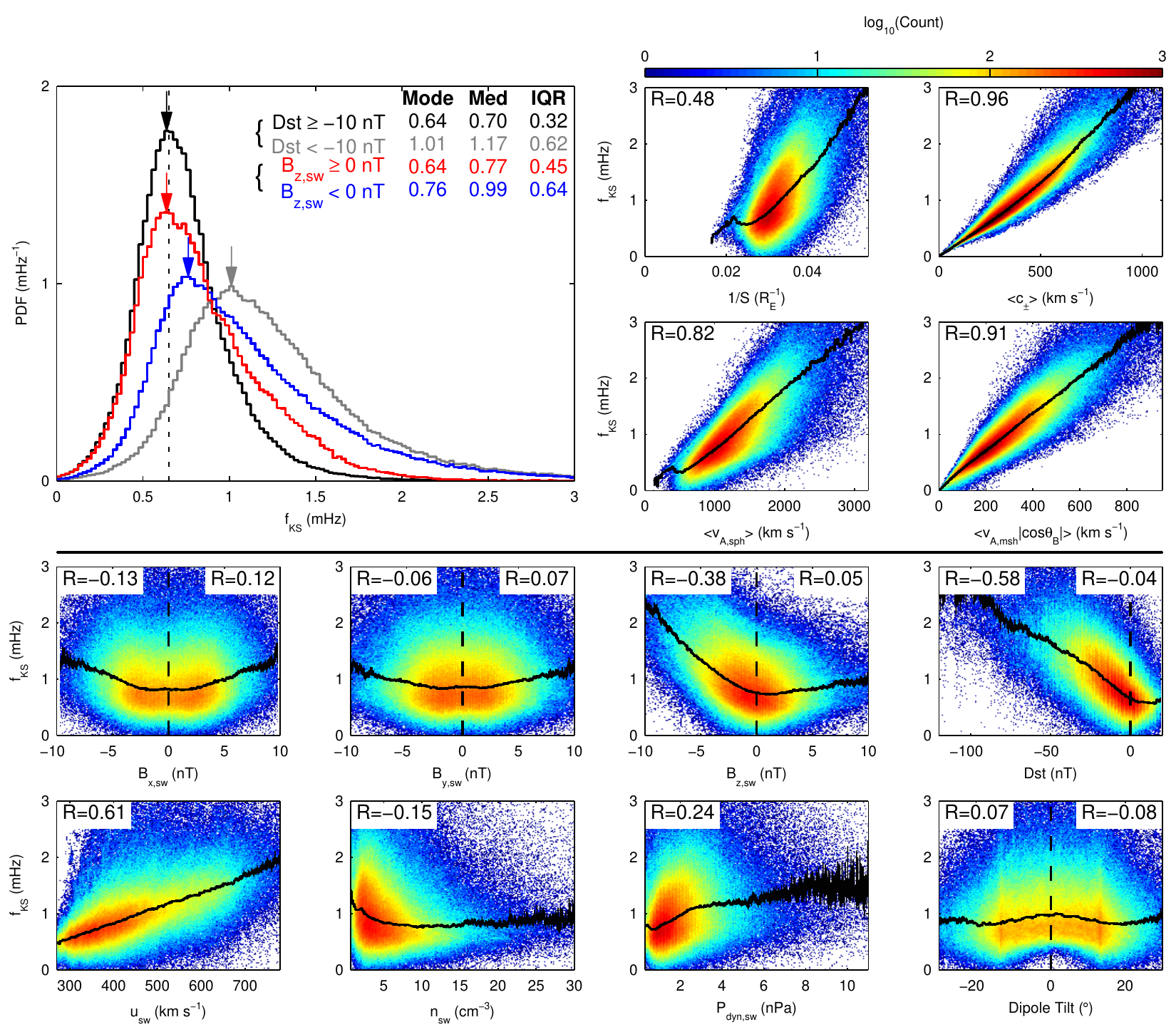}
\par\end{centering}

\protect\caption{Top Left: Distribution of calculated fundamental KS mode frequencies
during storm (grey) and non-storm (black) times and under northward
(red) and southward (blue) IMF respectively. The mode (indicated by
the arrows), median and interquartile range are also given for each.
The vertical dotted line is at 0.65~mHz. Top Right: Bivariate histograms
showing the dependence of the fundamental KS mode frequency on the
(reciprocal of the) field line length, average phase speed, average
magnetospheric Alfv\'{e}n speed and the average magnetosheath Alfv\'{e}n
speed projected along the geomagnetic field direction. The logarithmic
colour scale shows the number of datapoints in each bin and the black
line indicates the median frequency for each horizontal bin. Bottom:
Bivariate histograms in the same format showing the dependence of
the fundamental KS mode frequency on the input parameters - GSM x,
y and z components of the IMF, the $Dst$ index, solar wind speed,
density, dynamic pressure and the dipole tilt angle.\label{fig:distributions}}
\end{figure}

To ascertain whether the KS mode frequency is predominantly an average
phase speed or field line length effect, we plot bivariate histograms
of $f_{KS}$ against the reciprocal of the field line length $S\equiv\int ds$
and the average phase speed $\left\langle c_{\pm}\right\rangle \equiv2Sf_{KS}$
in Figure~\ref{fig:distributions}(top right). The average phase
speed was found to be typically $\sim$1.3 times larger than the phase
speed at the nose, used in previous frequency calculations \citep{plaschke09a,plaschke11,archer13c}.
Since the correlation coefficient is susceptible to skewness and outliers,
we use the correlation median estimator $R$ \citep{shevylakov87,falk98}
to assess the dependences. While both the field line length and average
phase speed do affect $f_{KS}$ as indicated by the medians (black
lines), the correlation shows the average phase speed dominates.

The average phase speed can be thought of as a mean weighted by the
time-of-flight of the surface waves, thus we introduce a generalised
weighted average given by

\begin{equation}
\left\langle a\right\rangle \equiv\left[\sum_{\pm}\int\frac{a}{c_{\pm}}ds\right]/\left[\sum_{\pm}\int\frac{ds}{c_{\pm}}\right]
\end{equation}
and determine which of the magnetospheric or magnetosheath Alfv\'{e}n
speeds (where only the component of the latter along the geomagnetic
field line is used) is most important in prescribing $f_{KS}$. The
bivariate histograms of these are also shown in Figure~\ref{fig:distributions}
(top right) revealing high correlation in both quantities, though
the magnetosheath Alfv\'{e}n speed correlates better hence KS mode
frequencies are slightly more dependent on magnetosheath properties
than those in the magnetosphere.

Figure~\ref{fig:distributions} (bottom) shows bivariate histograms
of $f_{KS}$ against the input parameters (spanning 0.5-99.5\%) to
our calculations. We find that the fundamental KS mode frequency depends
most strongly on the solar wind speed, disturbance storm time index
($Dst$) when negative and southward component of the IMF. All other
variables correlate poorly and the slope of the medians (black) are
small for the majority of the data, indicating little dependence.

We perform a multiple linear regression on the fundamental standing
surface wave frequency, where we normalise the input parameters $x_{i}$
by subtracting their median values and dividing by the interquartile
range i.e.

\begin{equation}
f_{KS}=a_{0}+\sum_{i}a_{i}\left[\frac{x_{i}-\mathrm{Med}\left(x_{i}\right)}{\mathrm{IQR}\left(x_{i}\right)}\right]\label{eq:regression}
\end{equation}
where $a_{0}$ is a constant. The coefficients $a_{i}$ of the regression
give a measure of the partial derivative of the frequency (keeping
all other variables fixed) to each variable whereas the previous correlations
relate to the total derivatives. The resulting regression coefficients
(as well as the normalisations) are shown in Table~\ref{tab:regression-results}.
While the residuals of this linear model are comparable to the spread
of the distribution of $f_{KS}$ and thus should not be used to estimate
the frequency, it does nonetheless reveal the relative dependences
of the inputs on the frequency. Again we see that southward IMF, the
solar wind speed and the $Dst$ index affect $f_{KS}$ the most, with
all other variables having order of magnitude smaller coefficients.

\subsection{Interpretation}

Here we interpet the physics governing the dependences discovered
in the previous section.

\subsubsection{$u_{sw}$ dependence}

\citet{archer13c} also discovered a dependence of expected standing
surface wave frequencies of the subsolar magnetopause on the solar
wind speed for 130 subsolar magnetosheath jet events, with this high
correlation being an effect of the phase speed at the nose (the assumed
constant phase speed over the entire field line). The phase speed
of a surface wave at the magnetopause nose is given by

\begin{subequations}\label{eq:usw_derivation}

\begin{center}
\begin{align}
c_{0} & =\sqrt{\frac{B_{msh,0}^{2}\cos^{2}\theta_{B}+B_{sph,0}^{2}}{\mu_{0}\left(\rho_{msh,0}+\rho_{sph,0}\right)}} & \begin{array}{c}
B_{sph,0}\gg B_{msh,0}\\
\rho_{msh,0}\gg\rho_{sph,0}
\end{array}\label{eq:usw_derivation1}\\
 & \simeq\sqrt{\frac{B_{sph,0}^{2}}{\mu_{0}\rho_{msh,0}}} & \begin{array}{c}
B_{sph,0}^{2}/2\mu_{0}=\rho_{sw}u_{sw}^{2}\\
\rho_{msh,0}\sim4.23\rho_{sw}
\end{array}\label{eq:usw_derivation3}\\
 & \simeq0.69u_{sw}\label{eq:usw_answer}
\end{align}

\par\end{center}

\end{subequations}demonstrating the approximate linear relation with
solar wind speed. A least-squares linear fit of the calculated nose
phase speed from our model results in a coefficient of $0.739$, consistent
with that from our simple derivation, and we find a correlation $R=0.88$.
Since $c_{0}$ also highly correlates with the average phase speed
($R=0.84$), this explains the strength of the determined relationship
of the frequency with solar wind speed ($0.88\times0.84\times0.91=0.67$
is close to the $R=0.61$ correlation of $f_{KS}$ with $u_{sw}$).

\subsubsection{$B_{z,sw}<0$ and $Dst$ dependences}

Here we consider both the southward component of the IMF and $Dst$
dependences on the fundamental KS mode frequency. These two variables
are not entirely independent, with large negative values of $Dst$
indicating storm times typically occurring during times of southward
IMF \citep{burton75}. Nonetheless, there are times (about 16\% of
all observations) when $Dst$ indicates storm times whereas the IMF
is northward. Similarly, 27\% of all times we find that the IMF is
southward yet $Dst$ indicates non-storm times. Since $Dst$ and $B_{z,sw}$
are both inputs to our model and are determined separately, with the
GSM z component of the IMF measured in the solar wind and $Dst$ determined
from ground magnetometer measurements, we have thus far treated them
independently in our statistical analyses.

\begin{figure}
\begin{centering}
\includegraphics{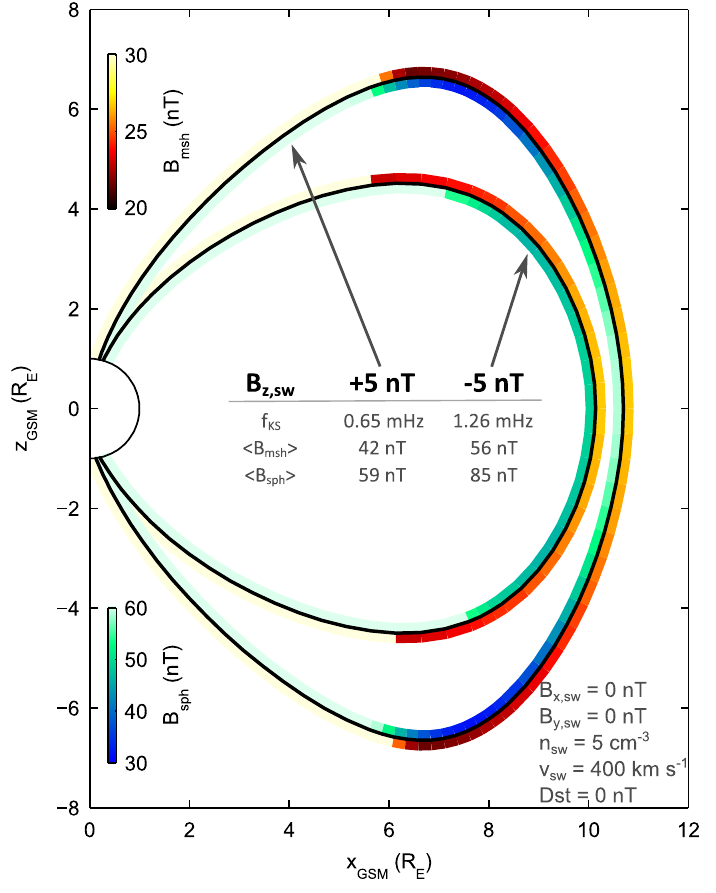}
\par\end{centering}

\protect\caption{Examples of T96 subsolar magnetopause field lines under northward
and southward IMF (black) with all other inputs kept constant. The
colours either side of the field line represent the magnetic field
strength on the magnetospheric and magnetosheath sides of the boundary,
from the T96 and KF94 models respectively.\label{fig:Bz}}
\end{figure}

We interpret the dependence of the fundamental standing surface wave
frequency on the southward component of the IMF as being due to the
T96 model's parameterisation of the erosion of the dayside magnetosphere
by reconnection. \citet{dungey61} described how reconnection leads
to shrinkage of the dayside magnetopause and a transport of magnetic
flux from the dayside to the nightside, resulting in equatorward motion
of the cusps \citep[e.g.][]{burch73,newell89}. Therefore, under southward
IMF the shape of the subsolar magnetopause field line is changed whereby
the standoff distance is shorter and the field line is less extended
in the GSM z direction. This can be seen in the examples shown in
Figure~\ref{fig:Bz} for both northward and southward IMF keeping
all other inputs constant. While the field line is shorter under southward
IMF, which will have the effect of increasing the fundamental frequency,
we have already established that the average phase speed is more important
in terms of the frequency. The typical reduction of the magnetospheric
and magnetosheath magnetic fields near the cusps won't be as large
for southward IMF since the field lines do not extend as far in the
z direction. This effect can clearly be seen in Figure~\ref{fig:Bz}.
The overall effect is that the average magnetic fields on both sides
of the boundary over the field line are increased compared to northward
IMF, thereby increasing the Alfv\'{e}n speeds (the average densities
are barely affected) resulting in higher frequencies.

The disturbance storm time index measures the intensity of the ring
current, whereby negative values mean that Earth's magnetic field
is weakened. The primary causes of geomagnetic storms are strong dawn-dusk
electric fields associated with the passage of southward IMF, with
reconnection providing the energy transfer between the IMF and the
magnetosphere \citep[e.g.][]{gonzalez94}. Therefore during storm
times, identified by strong negative $Dst$, the dayside magnetosphere
will again be eroded. Example calculations varying $Dst$ but keeping
all other inputs constant (not shown) are indeed very similar to those
varying only $B_{z,sw}$. We therefore interpret the effect of negative
$Dst$ on the fundamental KS mode frequency to also be due to the
T96 model's parameterisation of this erosion.

\section{Discussion}

\subsection{Validity of calculations\label{sub:Validity}}

Here we discuss the validity and accuracy of our KS mode fundamental
eigenfrequency calculations, in particular how known effects not captured
in this study may affect our results. These are summarised in Table
\ref{tab:approximations} where the median percentage differences
are given along with standard deviations.

Firstly we assess the dependence of the computed frequencies on the
specific magnetosheath and magnetospheric model quantities used in
this paper through sensitivity tests i.e. changing the models used
and their parameters \citep[c.f.][]{berube06,mccollough08}. The computed
frequencies are largely insensitive to the precise models used as
previously noted by \citet{archer13c} e.g. altering the magnetospheric
density by a factor of 2 affects $f_{KS}$ by less than 3\%, in contrast
to FLRs whose frequencies are highly dependent on the magnetospheric
mass density \citep[e.g.][]{waters96}. Overall we estimate the accuracy
of the computed frequencies due to the choice of models used to be
$\sim\pm10\%$.

The time-of-flight technique used in this paper essentially relies
on the WKB approximation to the solution of the full wave equation.
However, since the wavelengths in consideration for the fundamental
eigenfrequency are comparable to or larger than, for example, the
scale size of density and magnetic field variations along a field
line (e.g. Figure~\ref{fig:Example}) this is not strictly justified
\citep{singer91,schulz96,rankin06,kabin07}. This effect has been
shown to be small, but not negligible, for FLRs applied to similar
models to those here through numerically solving the full wave equation
in the model geometry \citep[e.g.][]{wild05}. Thus the use of time-of-flight
analysis in this study may affect the exact numerical solutions. It
can be shown by inserting the assumed WKB solution into a general
wave equation that, to the next order, a correction factor to the
time-of-flight integral in Equation~\ref{eq:tof} applies, given
by \citep[e.g.][]{kroemer94}

\begin{equation}
\sqrt{1+\frac{1}{2}\frac{k_{\mu}^{\prime\prime}}{k_{\mu}}-\frac{3}{4}\left(\frac{k_{\mu}^{\prime}}{k_{\mu}}\right)^{2}}
\end{equation}

where primes indicate the spatial derivative along the field line.
Calculating this correction factor reveals it alters our frequency
estimates by $+15\pm4\%$ i.e. a reasonably systematic effect on our
results.

On a similar note, the time-of-flight calculations were applied to
a local dispersion relation whereas the subsolar magnetopause field
line is clearly curved. \citet{singer91} derived a wave equation,
applied to standing Alfv\'{e}n waves, in a generalised magnetic field
geometry through the introduction of a geometry-dependent scale factor
$h_{\alpha}$, the relative normal distance between field lines given
some initial displacement vector (see also \citet{rankin06} and \citet{kabin07}).
It can be easily shown from the resulting wave equation that while
these spatially varying $h_{\alpha}$ factors affect the local amplitudes
and damping rates of waves, it has no effect on the phase (and thus
no effect on the wave frequency in a time-of-flight calculation).
Therefore our results are not changed by this consideration.

The surface wave dispersion relation used in this study (Equation~\ref{eq:dispersion})
assumes there is no azimuthal component to the wavevector. This is
justifiable because a surface wave with a significant azimuthal component
to its group velocity $\partial\omega/\partial\mathbf{k}$ will be
convected tailward down the flanks by the fast magnetosheath flow
(see Figure~\ref{fig:cartoon}). As discussed in the introduction,
standing surface waves are likely not possible away from the subsolar
magnetopause, though KS modes could perhaps be supported with some
azimuthal propagation. We therefore assess the change in the total
surface wave phase speed at the magnetopause nose by introducing a
small $k_{\phi}$, finding this has little effect ($-0.5\pm0.4$\%
for $k_{\phi}=0.1k_{\mu}$) on our results.

Our calculations also assume plasma incompressibility, however \citet{plaschke11}
showed that this isn't strictly valid at the magnetopause using typical
conditions. We assess the validity of the incompressibility assumption
here using the parameter
\begin{equation}
\frac{K^{2}}{k_{\mu}^{2}}\equiv\frac{\omega^{2}}{v_{A}^{2}+c_{s}^{2}\left[\omega^{2}-\left(\mathbf{k}\cdot\mathbf{v}_{A}\right)^{2}\right]/\omega^{2}}
\end{equation}
whereby incompressibility is valid if $\left|K^{2}/k_{\mu}^{2}\right|\ll1$
\citep{plaschke11}. Using the results of our analysis, this is estimated
at each point along the field line as

\begin{subequations}

\begin{align}
\frac{K_{msh}^{2}}{k_{\mu}^{2}} & =\frac{c_{\pm}^{2}\left(1\mp\frac{u_{msh}}{c_{\pm}}\right)^{2}}{v_{A,msh}^{2}+c_{s,msh}^{2}\left[\left(1\mp\frac{u_{msh}}{c_{\pm}}\right)^{2}-\frac{v_{A,msh}^{2}}{c_{\pm}^{2}}\cos^{2}\theta_{B}\right]/\left(1\mp\frac{u_{msh}}{c_{\pm}}\right)^{2}}\label{eq:incompressibility-msh}\\
\frac{K_{sph}^{2}}{k_{\mu}^{2}} & =\frac{c_{\pm}^{2}}{v_{A,sph}^{2}+c_{s,sph}^{2}\left[1-\frac{v_{A,sph}^{2}}{c_{\pm}^{2}}\right]}\label{eq:incompressibility-sph}
\end{align}

\end{subequations}

\begin{figure}
\begin{centering}
\includegraphics{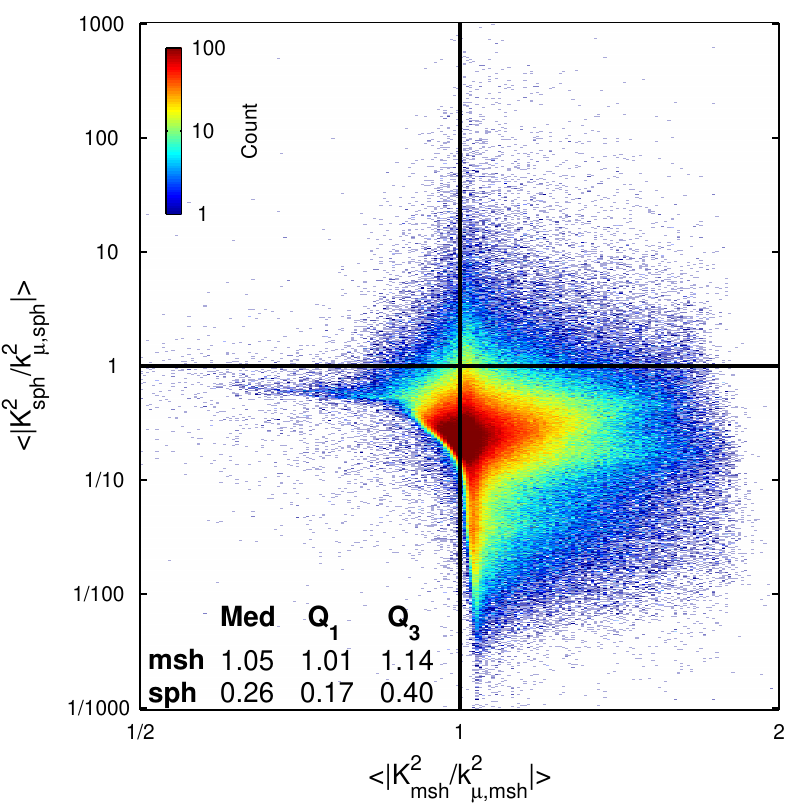}
\par\end{centering}

\protect\caption{Bivariate histogram of the average values of $\left|K^{2}/k_{\mu}^{2}\right|$
over the field lines, used to assess the incompressibility assumption,
in both the magnetosheath (horizontal axis) and magnetosphere (vertical
axis). The logarithmic colour scale indicates the number of datapoints
in each bin. The medians, first (25\%) and third (75\%) quartiles
are also indicated for both the magnetosheath and magnetosphere.\label{fig:Incompressibility}}
\end{figure}

The sound speed $c_{s}$ is computed in the magnetosphere by assuming
a plasma $\beta$ of 0.15 \citep[e.g.][]{phan94} and in the magnetosheath
by pressure balance i.e. the magnetosheath thermal pressure is given
by $\left(1+\beta_{sph}\right)\times$ the magnetic pressure of T96
minus the magnetic pressure of the KF94 draping model. Figure~\ref{fig:Incompressibility}
shows a bivariate histogram of the average values of $\left|K^{2}/k_{\mu}^{2}\right|$
over the field lines in both the magnetosheath (horizontal axis) and
magnetosphere (vertical axis), where the logarithmic colour scale
indicates the number of datapoints in each bin. The medians, first
(25\%) and third (75\%) quartiles are also indicated for both the
magnetosheath and magnetosphere. It is clear that the incompressibility
assumption $\left|K^{2}/k_{\mu}^{2}\right|\ll1$ is generally valid
in the magnetosphere but not in the magnetosheath. The full compressible
plasma dispersion relation is a 10th order polynomial for which no
general analytical solution exists, hence it would have to be solved
numerically at each point on the field line. This would result in
multiple solutions corresponding to different modes (e.g. S and F
modes \citep{pu83}) which would have to be carefully identified and
matched together at each point, making the calculation of KS mode
frequencies much more difficult. To assess the effect magnetosheath
plasma compressibility has on our results, we construct a new dispersion
relation applicable at the magnetopause nose where the compressible
plasma relation is included for the magnetosheath only. This yields
a quartic equation in the square of the surface wave phase speed which
is solved numerically and any positive real solutions (corresponding
to stable surface waves) are compared with the phase speed from Equation~\ref{eq:dispersion}.
We find that taking account of magnetosheath compressibility also
has a reasonably systematic effect on our results, adjusting them
by $-20\pm4\%$.

When the magnetic shear $\theta_{B}$ between the draped IMF and the
geomagnetic field is small, plasma depletion can occur resulting in
increased magnetic fields and reduced plasma densities on the magnetosheath
side of the boundary \citep{zwan76}. The width of this plasma depletion
layer (PDL) depends on the Alfv\'{e}nic Mach number of the bow shock
$M_{A}$, with large PDL's being possible under low $M_{A}$. Our
models do not include a PDL, which would serve to increase the magnetosheath
Alfv\'{e}n speed, thereby increasing the KS mode frequency from those
estimated here. Using the results of \citet{paschmann93}, we find
that plasma depletion should modify our calculations by $+60\pm20\%$.
However, this process should only be prevalent $\sim$1\% of the time
overall ($\theta_{B}<30^{\circ}$ and $M_{A}<8$), hence does not
significantly alter our statistical results.

Finally, if reconnection is occurring at any point along the subsolar
magnetopause field line, then the magnetosphere will be open and standing
surface waves will not be possible. We have not accounted for the
occurrence of reconnection in our distributions, though of course
this will not affect our results under northward IMF. Using its $\Delta\beta$-$\theta_{B}$
dependence \citep{swisdak10,phan13} we estimate reconnection was
allowed at most 64\% of the time under southward IMF. Note that this
is a necessary but not a sufficient condition in establishing whether
reconnection may be occurring at any time in our calculations. Nonetheless,
removing these times we find that the spread of our distributions
for storm times and under southward IMF are reduced by 0.1-0.2~mHz
and that the most likely frequency under southward IMF becomes 0.7~mHz.
These are relatively small changes to our results, which have mostly
concentrated on the unaffected northward IMF and non-storm times anyway.

Combining all of these effects on our calculations, we estimate that
the overall accuracy of the results presented here is $_{-19}^{+4}\%$.
Future modelling work into standing surface waves of the subsolar
magnetopause should attempt to incorporate some or all these effects
fully to give a more accurate description of the eigenmode. Many of
these considerations, however, will necessitate significantly more
computationally intensive calculations, hence may not be suitable
for such a large statistical database as that presented here. A number
of case studies should thus be modelled and the eigenfrequencies found
may then be compared with those presented here.

\subsection{Implications}

The results of our calculations provide a database of expected fundamental
frequencies of KS modes at the subsolar magnetopause, given the magnetospheric
and solar wind conditions at each time. The possible eigenfrequencies
of the subsolar magnetopause at any given time thus correspond to
integer multiples of this fundamental frequency $f_{KS}$ i.e. the
harmonics of the standing surface waves are $qf_{KS}$ where $q\in\mathbb{N}$.
Using the most likely frequency under both northward IMF and non-storm
times, this corresponds to \{0.6, 1.3, 1.9, 2.6, 3.2\ldots\}~mHz,
consistent with the reported discrete frequencies interpreted as indirect
evidence of KS modes \citep{plaschke09a,archer13c}.

\begin{figure}
\begin{centering}
\includegraphics{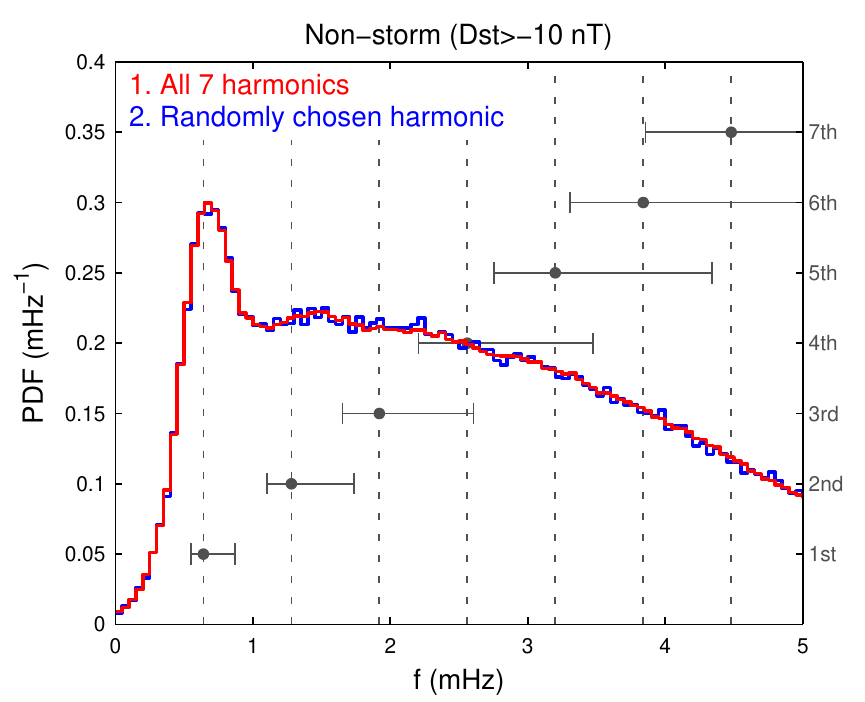}
\par\end{centering}

\protect\caption{Estimates of oscillation frequency occurrence distributions due to
KS mode harmonics during non-storm times, assuming either: 1. all
7 harmonics are present at all times (red) or 2. one randomly chosen
harmonic is present at each time (blue). The harmonics (integer multiples)
of the most likely fundamental frequency (c.f. Figure~\ref{fig:distributions}
top left) are shown as the vertical dotted lines, with the error bars
indicating their interquartile ranges.\label{fig:harmonics}}
\end{figure}

However, our distributions of $f_{KS}$ (Figure~\ref{fig:distributions}
top left) exhibit significant spread (of order $50\%$), which should
result in a large range of frequencies for the higher harmonics. Indeed,
the error bars in Figure~\ref{fig:harmonics} indicate the interquartile
ranges of the first 7 harmonics of KS modes from our non-storm time
calculations, revealing much overlap in frequency between the different
harmonics. We therefore wish to understand what effect the significant
spread in the fundamental frequency would have on observational statistics.
We use two different simple methods to approximate the distributions
of oscillation frequencies due to standing surface waves of the subsolar
magnetopause that could potentially be observed:
\begin{enumerate}
\item Assume that the first 7 harmonics are present at all times
\item Randomly choose just one of the first 7 harmonics at each time
\end{enumerate}
Both of these methods produce frequency distributions (shown in Figure~\ref{fig:harmonics})
whereby the most likely fundamental frequency of 0.64~mHz shows a
prominant peak, whereas no further significant peaks are found i.e.
the overtones of the most likely fundamental eigenfrequency are not
apparent in these occurence distributions. Therefore, while our results
suggest a most likely set of KS mode eigenfrequencies may exist, they
also predict that occurrence distributions of Pc5 frequencies due
to KS modes would actually result in a continuum. This is in contrast
to the results of \citet{plaschke09} showing prominant observed oscillation
frequencies of the subsolar magnetopause which were then attributed
to KS modes. There are, however, a couple of possible explanations
why discrete frequencies due to subsolar magnetopause eigenmodes may
occur in such statistical studies. If statistics are poor and therefore
do not cover the full range of solar wind conditions or there is some
unconscious selection bias towards certain solar wind conditions,
then the sample distribution of frequencies may not be representative
of the full distribution. It is thus possible that distinct peaks
may emerge under these circumstances. An alternative explanation comes
down to the presence or not of a suitable driver. Our distributions
make no predictions on whether standing surface waves may actually
be present at the subsolar magnetopause at any given time. Such a
statement would require consideration of some particular driver for
this eigenmode such as solar wind pressure pulses or localised magnetosheath
jets \citep{archer13c}. If such a driver existed at some time, then
the KS modes could be excited at one of the harmonics of our calculated
fundamental frequency $f_{KS}$ or a combination thereof. However,
such drivers may preferentially occur under certain solar wind conditions.
Indeed, \citet{plaschke09} found that subsolar magnetopause oscillations
tended to occur under low cone-angle IMF, conditions for which magnetosheath
jets predominantly occur \citep{archer13,plaschke13}.

Since magnetopause surface waves are evanescent in the magnetosphere,
their signatures should decay exponentially with distance from the
boundary. Assuming incompressibility, typically valid on the magnetospheric
side of the boundary as shown in section \ref{sub:Validity}, we have
$k^{2}\equiv k_{\nu}^{2}+k_{\phi}^{2}+k_{\mu}^{2}=0$ \citep{plaschke11}
and thus in our calculations the magnitude of the (imaginary) radial
component of the wavevector$\left|k_{\nu}\right|=\left|k_{\mu}\right|$.
The evanescent length scale at the magnetopause nose is therefore
given by $\left|k_{\nu}\right|^{-1}=c_{0}/(2\pi qf_{KS})$, where
$q$ is again the harmonic number of the KS mode. We calculate this
length scale for the most likely KS mode harmonics during non-storm
times (with frequencies given earlier) yielding \{8.5, 4.2, 2.8, 2.14,
1.7\ldots\}~$\mathrm{R_{E}}$. Thus while the higher harmonics are
somewhat confined to the vicinity of the subsolar magnetopause, the
first few harmonics could potentially be detected at, for example,
geostationary orbit. Previous work has shown that discrete Pc5 oscillation
frequencies in the subsolar magnetosphere at geostationary orbit can
be explained as directly driven waves 54\% of the time that periodic
density structures exist in the solar wind \citep{viall09}. However,
discrete compressional Pc5 oscillations have also been observed due
to broadband magnetosheath jets in the absence of such monochromatic
solar wind dynamic pressure fluctuations and these have been interpreted
as KS modes \citep{archer13c}. While the results presented here suggest
this interpretation is indeed plausible, the current indirect observational
evidence for KS modes should be carefully reassessed in the context
of this study to ascertain whether those observed frequencies can
indeed be explained as due to this eigenmode of the subsolar magnetopause.
While KS modes have been proposed as a potential source of discrete
field line resonances in general \citep{plaschke09a}, whether such
coupling can occur at some location inside the subsolar magnetosphere
will be highly dependent on the FLR frequency profile within the KS
mode's spatially confined extent. It is unclear at present how often
such coupling may occur, which is beyond the scope of this study but
could form the basis of future work. However, we limit our discussions,
analysis and implications of standing subsolar surface waves of the
magnetopause to the subsolar region only and make no claim that these
oscillations can directly excite discrete FLRs in the magnetospheric
flanks, such as those reported by \citet{samson91,samson92}.

\section{Conclusions}

In this paper we have presented the first estimates of the distribution
of standing surface wave frequencies at the subsolar magnetopause
using the time-of-flight technique \citep[e.g.][]{wild05} applied
to combined models of the magnetosphere and magnetosheath. We find
that the most likely frequency during non-storm times or under northward
IMF is $0.64_{-0.12}^{+0.03}$~mHz, consistent with the fundamental
frequency of $\sim$0.65~mHz proposed by \citet{plaschke09a} from
a simple estimate using typical conditions and the approximate regularity
of observed oscillation periods of the subsolar boundary. However,
the distributions exhibit a large amount of spread (of order $\pm$0.3~mHz
or 50\%), sufficient that the overtones of the most likely frequency
are not apparent when constructing distributions of the KS mode harmonics
over the full range of solar wind conditions.

We find that the KS mode frequencies principally depend on the solar
wind speed, in agreement with \citet{archer13c}, as well as the disturbance
storm time ($Dst$) index and the southward component of the IMF.
We have ascertained the physical reasons for these dependences, with
the latter two being due to the erosion of the dayside magnetosphere
by reconnection and the solar wind speed dependence a result of the
phase speed of surface waves at the magnetopause nose (in turn proportional
to the average phase speed). Finally, we present that the possible
occurrence of KS modes (reconnection notwithstanding) is primarily
controlled by the dipole tilt angle, since the reversal of either
the parallel or antiparallel propagating surface waves by the magnetosheath
flow is predominantly a geometrical effect.

Future work will compare the magnetospheric ULF wave activity in spacecraft
and ground magnetometer data with our database of expected KS mode
frequencies. It is clear though that care must be taken for example
in identifying harmonics in the observations and accounting for all
possible ULF wave drivers and modes for each event. By doing this
it may be possible to not only validate our model calculations, but
provide further evidence for the possible existence of eigenmodes
of the subsolar magnetopause.
\begin{acknowledgments}
M. O. Archer is thankful for funding through STFC grant ST/I505713/1.
The OMNI data was obtained from the NASA/GSFC OMNIWeb interface at
http://omniweb.gsfc.nasa.gov. We thank the editor for their helpful
suggestions during the review process.
\end{acknowledgments}


\begin{thebibliography}{63}
\providecommand{\natexlab}[1]{#1}
\expandafter\ifx\csname urlstyle\endcsname\relax
  \providecommand{\doi}[1]{doi:\discretionary{}{}{}#1}\else
  \providecommand{\doi}{doi:\discretionary{}{}{}\begingroup
  \urlstyle{rm}\Url}\fi

\bibitem[{\textit{Archer and Horbury}(2013)}]{archer13}
Archer, M.~O., and T.~S. Horbury, Magnetosheath dynamic pressure enhancements:
  {O}ccurrence and typical properties, \textit{Ann. Geophys.}, \textit{31},
  319--331, \doi{10.5194/angeo-31-319-2013}, 2013.

\bibitem[{\textit{Archer et~al.}(2013a)\textit{Archer, Horbury, Eastwood,
  Weygand, and Yeoman}}]{archer13b}
Archer, M.~O., T.~S. Horbury, J.~P. Eastwood, J.~M. Weygand, and T.~K. Yeoman,
  Magnetospheric response to magnetosheath pressure pulses: {A} low pass filter
  effect, \textit{J. Geophys. Res.}, \textit{118}, 5454--5466,
  \doi{10.1002/jgra.50519}, 2013a.

\bibitem[{\textit{Archer et~al.}(2013b)\textit{Archer, Hartinger, and
  Horbury}}]{archer13c}
Archer, M.~O., M.~D. Hartinger, and T.~S. Horbury, Magnetospheric ``magic''
  frequencies as magnetopause surface eigenmodes, \textit{Geophys. Res. Lett.},
  \textit{40}, 5003--5008, \doi{10.1002/grl.50979}, 2013b.

\bibitem[{\textit{Baker et~al.}(2003)\textit{Baker, Donovan, and
  Jackel}}]{baker03}
Baker, G.~J., E.~F. Donovan, and B.~J. Jackel, A comprehensive survey of
  auroral latitude {P}c5 pulsation characteristics, \textit{J. Geophys. Res.},
  \textit{108}, SMP 11--1, \doi{10.1029/2002JA009801}, 2003.

\bibitem[{\textit{Berube et~al.}(2006)\textit{Berube, Moldwin, and
  Ahn}}]{berube06}
Berube, D., M.~B. Moldwin, and M.~Ahn, Computing magnetospheric mass density
  from field line resonances in a realistic magnetic field geometry, \textit{J.
  Geophys. Res}, \textit{111}, A08,206, \doi{10.1029/2005JA011450}, 2006.

\bibitem[{\textit{Burch}(1973)}]{burch73}
Burch, J.~L., Rate of erosion of dayside magnetic flux based on a quantitative
  study of the dependence of polar cusp latitude on the interplanetary magnetic
  field, \textit{Radio Sci.}, \textit{8}, 955--961,
  \doi{10.1029/RS008i011p00955}, 1973.

\bibitem[{\textit{Burton et~al.}(1975)\textit{Burton, Mc{P}herron, and
  Russell}}]{burton75}
Burton, R.~K., R.~L. Mc{P}herron, and C.~T. Russell, An empirical relationship
  between interplanetary conditions and dst, \textit{J. Geophys. Res.},
  \textit{80}, 4204--4214, \doi{10.1029/JA080i031p04204}, 1975.

\bibitem[{\textit{Chen and Hasegawa}(1974)}]{chen74}
Chen, L., and A.~Hasegawa, A theory of long-period magnetic pulsations: 2.
  impulse excitation of surface eigenmode, \textit{J. Geophys. Res.},
  \textit{79}, 1033--1037, \doi{10.1029/JA079i007p01033}, 1974.

\bibitem[{\textit{Chisham and Orr}(1997)}]{chisham97}
Chisham, G., and D.~Orr, A statistical study of the local time asymmetry of
  {P}c5 {ULF} wave characteristics observed at midlatitudes by {SAMNET},
  \textit{J. Geophys. Res.}, \textit{102}, 24,339--24,350,
  \doi{10.1029/97JA01801}, 1997.

\bibitem[{\textit{Claudepierre et~al.}(2013)}]{claudepierre13}
Claudepierre, S.~G., et~al., {V}an {A}llen {P}robes observation of localized
  drift resonance between poloidal mode ultra-low frequency waves and 60 {keV}
  electrons, \textit{Geophys. Res. Lett.}, \textit{40}, 4491--4497,
  \doi{10.1002/grl.50901}, 2013.

\bibitem[{\textit{De~Keyser et~al.}(2005)\textit{De~Keyser, Dunlop, Owen,
  Sonnerup, Haaland, Vaivads, Paschmann, Lundin, and Rezeau}}]{dekeyser05}
De~Keyser, J., M.~W. Dunlop, C.~J. Owen, B.~U.~. Sonnerup, S.~E. Haaland,
  A.~Vaivads, G.~Paschmann, R.~Lundin, and L.~Rezeau, Magnetopause and boundary
  layer, \textit{Space Science Reviews}, \textit{118}, 231--320,
  \doi{10.1007/s11214-005-3834-1}, 2005.

\bibitem[{\textit{Denton et~al.}(2002)\textit{Denton, Goldstein, Menietti, and
  Young}}]{denton02}
Denton, R.~E., J.~Goldstein, J.~D. Menietti, and S.~L. Young, Magnetospheric
  electron density model inferred from {P}olar plasma wave data, \textit{J.
  Geophys. Res.}, \textit{107}, SMP 25--1 -- SMP 25--8,
  \doi{10.1029/2001JA009136}, 2002.

\bibitem[{\textit{Dungey}(1961)}]{dungey61}
Dungey, J.~W., Interplanetary magnetic field and the auroral zones,
  \textit{Phys. Rev. Lett.}, \textit{6}, 47--48,
  \doi{10.1103/PhysRevLett.6.47}, 1961.

\bibitem[{\textit{Falk}(1998)}]{falk98}
Falk, M., A note on the co-median for elliptical distributions, \textit{J.
  Multivar. Analysis}, \textit{67}, 306--317, \doi{10.1006/jmva.1998.1775},
  1998.

\bibitem[{\textit{Fenrich et~al.}(1995)\textit{Fenrich, Samson, Sofko, and
  Greenwald}}]{fenrich95}
Fenrich, F.~M., J.~C. Samson, G.~Sofko, and R.~A. Greenwald, {ULF} high- and
  low-m field line resonances observed with the {S}uper {D}ual {A}uroral
  {R}adar {N}etwork, \textit{J. Geophys. Res.}, \textit{100}, 21,535--21,547,
  \doi{10.1029/95JA02024}, 1995.

\bibitem[{\textit{Francia and Villante}(1997)}]{francia97}
Francia, P., and U.~Villante, Some evidence of ground power enhancements at
  frequencies of global magnetospheric modes at low latitude, \textit{Ann.
  Geophys.}, \textit{15}, 17--23, \doi{10.1007/s00585-997-0017-2}, 1997.

\bibitem[{\textit{Gonzalez et~al.}(1994)\textit{Gonzalez, Joselyn, Kamide,
  Kroehl, Rostoker, Tsurutani, and Vasyliunas}}]{gonzalez94}
Gonzalez, W.~D., J.~A. Joselyn, Y.~Kamide, H.~W. Kroehl, G.~Rostoker, B.~T.
  Tsurutani, and V.~M. Vasyliunas, What is a geomagnetic storm?, \textit{J.
  Geophys. Res.}, \textit{99}, 5771--5792, \doi{10.1029/93JA02867}, 1994.

\bibitem[{\textit{Johnson et~al.}(2014)\textit{Johnson, Wing, and
  Delamere}}]{johnson14}
Johnson, J.~R., S.~Wing, and P.~A. Delamere, Kelvin helmholtz instability in
  planetary magnetospheres, \textit{Space Sci. Rev.}, \textit{184}, 1--31,
  \doi{10.1007/s11214-014-0085-z}, 2014.

\bibitem[{\textit{Kabin et~al.}(2007)\textit{Kabin, Rankin, Waters, Marchand,
  Donovan, and Samson}}]{kabin07}
Kabin, K., R.~Rankin, C.~L. Waters, R.~Marchand, E.~F. Donovan, and J.~C.
  Samson, Different eigenproblem models for field line resonances in cold
  plasma: Effect on magnetospheric density estimates, \textit{Planet. Space
  Sci.}, \textit{55}, 820--828, \doi{10.1016/j.pss.2006.03.014}, 2007.

\bibitem[{\textit{Kepko and Spence}(2003)}]{kepko03}
Kepko, L., and H.~E. Spence, Observations of discrete, global magnetospheric
  oscillations directly driven by solar wind density variations, \textit{J.
  Geophys. Res.}, \textit{108}, SMP21--1 -- SMP21--12,
  \doi{10.1029/2002JA009676}, 2003.

\bibitem[{\textit{Kepko et~al.}(2002)\textit{Kepko, Spence, and
  Singer}}]{kepko02}
Kepko, L., H.~E. Spence, and H.~J. Singer, {ULF} waves in the solar wind as
  direct drivers of magnetospheric pulsations, \textit{Geophys. Res. Lett.},
  \textit{29}, 39--1 -- 39--4, \doi{10.1029/2001GL014405}, 2002.

\bibitem[{\textit{Kivelson and Southwood}(1985)}]{kivelson85}
Kivelson, M.~G., and D.~J. Southwood, Resonant {ULF} waves: a new
  interpretation, \textit{Geophys. Res. Lett.}, \textit{12}, 49--52,
  \doi{10.1029/GL012i001p00049}, 1985.

\bibitem[{\textit{Kobel and Fl\"{u}ckiger}(1994)}]{KF94}
Kobel, E., and E.~O. Fl\"{u}ckiger, A model of the steady state magnetic field
  in the magnetosheath, \textit{J. Geophys. Res.}, \textit{99}, 23,617--23,622,
  \doi{10.1029/94JA01778}, 1994.

\bibitem[{\textit{Kokubun}(2013)}]{kokubun13}
Kokubun, S., {ULF} waves in the outer magnetosphere: {G}eotail observation 1
  transverse waves, \textit{Earth Planets Space}, \textit{65}, 411--433,
  \doi{10.5047/eps.2012.12.013}, 2013.

\bibitem[{\textit{Kroemer}(1994)}]{kroemer94}
Kroemer, H., \textit{Quantum Mechanics for Engineering, Materials Science and
  Applied Physics}, Prentice Hall, 1994.

\bibitem[{\textit{Kruskal and Schwartzschild}(1954)}]{kruskal54}
Kruskal, M., and M.~Schwartzschild, Some instabilities of a completely ionized
  plasma, \textit{Proc. R. Soc. London, Ser. A.}, \textit{223}, 348--360,
  \doi{10.1098/rspa.1954.0120}, 1954.

\bibitem[{\textit{Landau and Lifshitz}(1959)}]{landau59}
Landau, L., and E.~Lifshitz, \textit{Fluid Mechanics}, Pergamon Press, 1959.

\bibitem[{\textit{Lavraud et~al.}(2004)\textit{Lavraud, Fedorov, Budnik,
  Grigoriev, Cargill, Dunlop, Rème, Dandouras, and Balogh}}]{lavraud04}
Lavraud, B., A.~Fedorov, E.~Budnik, A.~Grigoriev, P.~J. Cargill, M.~W. Dunlop,
  H.~Rème, I.~Dandouras, and A.~Balogh, Cluster survey of the high-altitude
  cusp properties: a three-year statistical study, \textit{Ann. Geophys.},
  \textit{22}, 3009--3019, \doi{10.5194/angeo-22-3009-2004}, 2004.

\bibitem[{\textit{Lee}(1996)}]{lee96}
Lee, D.-H., Dynamics of {MHD} wave propagation in the low-latitude
  magnetosphere, \textit{J. Geophys. Res.}, \textit{101}, 15,371--15,386,
  \doi{10.1029/96JA00608}, 1996.

\bibitem[{\textit{Lee and Lysak}(1989)}]{lee89}
Lee, D.-H., and R.~L. Lysak, Magnetospheric {ULF} wave coupling in the dipole
  model: The impulsive excitation, \textit{J. Geophys. Res.}, \textit{94},
  17,097--17,103, \doi{10.1029/JA094iA12p17097}, 1989.

\bibitem[{\textit{Mann et~al.}(2013)}]{mann13}
Mann, I.~R., et~al., Discovery of the action of a geophysical synchrotron in
  the {E}arth's {V}an {A}llen radiation belts, \textit{Nature Commun.},
  \textit{4}, 2795, \doi{10.1038/ncomms3795}, 2013.

\bibitem[{\textit{McCollough et~al.}(2008)\textit{McCollough, Gannon, Baker,
  and Gehmeyr}}]{mccollough08}
McCollough, J.~P., J.~L. Gannon, D.~N. Baker, and M.~Gehmeyr, A statistical
  comparison of commonly used external magnetic field models, \textit{Space
  Weather}, \textit{6}, S10,001, \doi{10.1029/2008SW000391}, 2008.

\bibitem[{\textit{Newell et~al.}(1989)\textit{Newell, Meng, and
  Sibeck}}]{newell89}
Newell, P.~T., C.-I. Meng, and D.~G. Sibeck, Some low-altitude cusp
  dependencies on the interplanetary magnetic field, \textit{J. Geophys. Res.},
  \textit{94}, 8921--8927, \doi{10.1029/JA094iA07p08921}, 1989.

\bibitem[{\textit{Paschmann et~al.}(1993)\textit{Paschmann, Baumjohann,
  Sckopke, Phan, and L\"{u}hr}}]{paschmann93}
Paschmann, G., W.~Baumjohann, N.~Sckopke, T.~D. Phan, and H.~L\"{u}hr,
  Structure of the dayside magnetopause for low magnetic shear, \textit{J.
  Geophys. Res.}, \textit{98}, 13,409--13,422, \doi{10.1029/93JA00646}, 1993.

\bibitem[{\textit{Pasman and Shevlyakov}(1987)}]{shevylakov87}
Pasman, V., and G.~L. Shevlyakov, Robust estimation of a correlation
  coefficient, \textit{Avtomat. Telemekh.}, \textit{3}, 70--80, 1987.

\bibitem[{\textit{Phan et~al.}(1994)\textit{Phan, Paschmann, Baumjohann,
  Sckopke, and L\"{u}hr}}]{phan94}
Phan, T.~D., G.~Paschmann, W.~Baumjohann, N.~Sckopke, and H.~L\"{u}hr, The
  magnetosheath region adjacent to the dayside magnetopause: {AMPTE}/{IRM}
  observations, \textit{J. Geophys. Res.}, \textit{99}, 121--141,
  \doi{10.1029/93JA02444}, 1994.

\bibitem[{\textit{Phan et~al.}(2013)\textit{Phan, Paschmann, Gosling, Oieroset,
  Fujimoto, Drake, and Angelopoulos}}]{phan13}
Phan, T.~D., G.~Paschmann, J.~T. Gosling, M.~Oieroset, M.~Fujimoto, J.~F.
  Drake, and V.~Angelopoulos, The dependence of magnetic reconnection on plasma
  $\beta$ and magnetic shear: {E}vidence from magnetopause observations,
  \textit{Geophys. Res. Lett.}, \textit{40}, 11--16,
  \doi{10.1029/2012GL054528}, 2013.

\bibitem[{\textit{Plaschke and Glassmeier}(2011)}]{plaschke11}
Plaschke, F., and K.~H. Glassmeier, Properties of standing
  kruskal-schwarzschild-modes at the magnetopause, \textit{Ann. Geophys.},
  \textit{29}, 1793--1807, \doi{10.5194/angeo-29-1793-2011}, 2011.

\bibitem[{\textit{Plaschke et~al.}(2009a)\textit{Plaschke, Glassmeier, Auster,
  Constantinescu, Magnes, Angelopoulos, Sibeck, and McFadden}}]{plaschke09a}
Plaschke, F., K.-H. Glassmeier, H.~U. Auster, O.~D. Constantinescu, W.~Magnes,
  V.~Angelopoulos, D.~G. Sibeck, and J.~P. McFadden, Standing alfvén waves at
  the magnetopause, \textit{Geophys. Res. Lett.}, \textit{36}, L02,104,
  \doi{10.1029/2008GL036411}, 2009a.

\bibitem[{\textit{Plaschke et~al.}(2009b)\textit{Plaschke, Glassmeier, Sibeck,
  Auster, Constantinescu, Angelopoulos, and Magnes}}]{plaschke09}
Plaschke, F., K.-H. Glassmeier, D.~G. Sibeck, H.~U. Auster, O.~D.
  Constantinescu, V.~Angelopoulos, and W.~Magnes, Magnetopause surface
  oscillation frequencies at different solar wind conditions, \textit{Ann.
  Geophys.}, \textit{27}, 4521--4532, \doi{10.5194/angeo-27-4521-2009}, 2009b.

\bibitem[{\textit{Plaschke et~al.}(2013)\textit{Plaschke, Hietala, and
  Angelopoulos}}]{plaschke13}
Plaschke, F., H.~Hietala, and V.~Angelopoulos, Anti-sunward high-speed jets in
  the subsolar magnetosheath, \textit{Ann. Geophys.}, \textit{31}, 1877--1889,
  \doi{10.5194/angeo-31-1877-2013}, 2013.

\bibitem[{\textit{Prikryl et~al.}(1998)\textit{Prikryl, Greenwald, Sofko,
  Villain, Ziesolleck, and Friis-Christensen}}]{prikryl98}
Prikryl, P., R.~A. Greenwald, G.~J. Sofko, J.~P. Villain, C.~W.~S. Ziesolleck,
  and E.~Friis-Christensen, Solar-wind-driven pulsed magnetic reconnection at
  the dayside magnetopause, pc5 compressional oscillations, and field line
  resonance, \textit{J. Geophys. Res}, \textit{103}, 17,307--17,322,
  \doi{10.1029/97JA03595}, 1998.

\bibitem[{\textit{Prikryl et~al.}(1999)\textit{Prikryl, MacDougall, Grant,
  Steele, Sofko, and Greenwald}}]{prikryl99}
Prikryl, P., J.~W. MacDougall, I.~F. Grant, D.~P. Steele, G.~J. Sofko, and
  R.~A. Greenwald, Observations of polar patches generated by solar wind
  alfv\'{e}n wave coupling to the dayside magnetosphere, \textit{Ann.
  Geophys.}, \textit{17}, 463--489, \doi{10.1007/s00585-999-0463-0}, 1999.

\bibitem[{\textit{Pu and Kivelson}(1983)}]{pu83}
Pu, Z.-Y., and M.~G. Kivelson, Kelvin-{H}elmholtz {I}nstability at the
  magnetopause: {S}olution for compressible plasmas, \textit{J. Geophys. Res.},
  \textit{88}, 841--852, \doi{10.1029/JA088iA02p00841}, 1983.

\bibitem[{\textit{Rae et~al.}(2012)\textit{Rae, Mann, Murphy, Ozeke, Milling,
  Chan, Elkington, and Honary}}]{rae12}
Rae, I.~J., I.~R. Mann, K.~R. Murphy, L.~G. Ozeke, D.~K. Milling, A.~A. Chan,
  S.~R. Elkington, and F.~Honary, Ground-based magnetometer determination of in
  situ {P}c4-5 {ULF} electric field wave spaectra as a function of solar wind
  speed, \textit{J. Geophys. Res.}, \textit{117}, A04,221,
  \doi{10.1029/2011JA017335}, 2012.

\bibitem[{\textit{Rankin et~al.}(2006)\textit{Rankin, Kabin, and
  Marchand}}]{rankin06}
Rankin, R., K.~Kabin, and R.~Marchand, Alfv\'{e}nic field line resonances in
  arbitrary magnetic field topology, \textit{Adv. Space Res.}, \textit{38},
  1720--1729, \doi{10.1016/j.asr.2005.09.034}, 2006.

\bibitem[{\textit{Russell}(2000)}]{russell00}
Russell, C.~T., The polar cusp, \textit{Adv. Space Res.}, \textit{25},
  1413--1424, \doi{10.1016/S0273-1177(99)00653-5}, 2000.

\bibitem[{\textit{Samson et~al.}(1991)\textit{Samson, Greenwald, Ruohoniemi,
  Hughes, and Wallis}}]{samson91}
Samson, J.~C., R.~A. Greenwald, J.~M. Ruohoniemi, T.~J. Hughes, and D.~D.
  Wallis, Magnetometer and radar observations of magnetohydrodynamic cavity
  modes in the {E}arth's magnetosphere, \textit{Can. J. Phys.}, \textit{69},
  929--937, \doi{10.1139/p91-147}, 1991.

\bibitem[{\textit{Samson et~al.}(1992)\textit{Samson, Harrold, Ruohoniemi,
  Greenwald, and Walker}}]{samson92}
Samson, J.~C., B.~G. Harrold, J.~M. Ruohoniemi, R.~A. Greenwald, and A.~D.~M.
  Walker, Field line resonances associated with {MHD} waveguides in the
  magnetosphere, \textit{Geophys. Res. Lett.}, \textit{19}, 441--444,
  \doi{10.1029/92GL00116}, 1992.

\bibitem[{\textit{Schulz}(1996)}]{schulz96}
Schulz, M., Eigenfrequencies of geomagnetic field lines and implications for
  plasma-density modeling, \textit{J. Geophys. Res.}, \textit{101},
  17,385--17,397, \doi{10.1029/95JA03727}, 1996.

\bibitem[{\textit{Shue et~al.}(1998)}]{shue98}
Shue, J.-H., et~al., Magnetopause location under extreme solar wind conditions,
  \textit{J. Geophys. Res.}, \textit{103}, 17,691--17,700,
  \doi{10.1029/98JA01103}, 1998.

\bibitem[{\textit{Singer et~al.}(1981)\textit{Singer, Southwood, Walker, and
  Kivelson}}]{singer91}
Singer, H.~J., D.~J. Southwood, R.~J. Walker, and M.~G. Kivelson, Alfv\'{e}n
  wave resonances in a realistic magnetospheric magnetic field geometry,
  \textit{J. Geophys. Res.}, \textit{86}, 4589--4596,
  \doi{10.1029/JA086iA06p04589}, 1981.

\bibitem[{\textit{Southwood}(1974)}]{southwood74}
Southwood, D.~J., Some features of field line resonances in the magnetosphere,
  \textit{Planet. Space Sci.}, \textit{22}, 483--491,
  \doi{10.1016/0032-0633(74)90078-6}, 1974.

\bibitem[{\textit{Spreiter et~al.}(1966)\textit{Spreiter, Summers, and
  Alksne}}]{spreiter66}
Spreiter, J.~R., A.~L. Summers, and A.~Y. Alksne, Hydromagnetic flow around the
  magnetosphere, \textit{Planet. Space Sci.}, \textit{14}, 223--250,
  \doi{10.1016/0032-0633(66)90124-3}, 1966.

\bibitem[{\textit{Swisdak et~al.}(2010)\textit{Swisdak, Opher, Drake, and
  {Alouani Bibi}}}]{swisdak10}
Swisdak, M., M.~Opher, J.~F. Drake, and F.~{Alouani Bibi}, The vector direction
  of the interstellar magnetic field outside the heliosphere, \textit{Ap. J.},
  \textit{710}, 1769--1775, \doi{10.1088/0004-637X/710/2/1769}, 2010.

\bibitem[{\textit{Tsyganenko}(1995)}]{t95}
Tsyganenko, N.~A., Modeling the earth's magnetospheric magnetic field confined
  within a realistic magnetopause, \textit{J. Geophys. Res.}, \textit{100},
  5599--5612, \doi{10.1029/94JA03193}, 1995.

\bibitem[{\textit{Tsyganenko and Stern}(1996)}]{t96}
Tsyganenko, N.~A., and D.~P. Stern, Modeling the global magnetic field of the
  large-scale {B}irkeland current systems, \textit{J. Geophys. Res.},
  \textit{101}, 27,187--27,198, \doi{10.1029/96JA02735}, 1996.

\bibitem[{\textit{Viall et~al.}(2008)\textit{Viall, Kepko, and
  Spence}}]{viall08}
Viall, N.~M., L.~Kepko, and H.~E. Spence, Inherent length-scales of periodic
  solar wind number density structures, \textit{J. Geophys. Res.},
  \textit{113}, A07,101, \doi{10.1029/2007JA012881}, 2008.

\bibitem[{\textit{Viall et~al.}(2009)\textit{Viall, Kepko, and
  Spence}}]{viall09}
Viall, N.~M., L.~Kepko, and H.~E. Spence, Relative occurrence rates and
  connection of discrete frequency oscillations in the solar wind density and
  dayside magnetosphere, \textit{J. Geophys. Res.}, \textit{114}, A01,201,
  \doi{10.1029/2008JA013334}, 2009.

\bibitem[{\textit{Villante et~al.}(2001)\textit{Villante, Francia, and
  Lepidi}}]{villante01}
Villante, U., P.~Francia, and S.~Lepidi, {P}c5 geomagnetic field fluctuations
  at discrete frequencies at a low latitude station, \textit{Ann. Geophys.},
  \textit{19}, 321--325, \doi{10.5194/angeo-19-321-2001}, 2001.

\bibitem[{\textit{Waters et~al.}(1996)\textit{Waters, Samson, and
  Donovan}}]{waters96}
Waters, C.~L., J.~C. Samson, and E.~F. Donovan, Variation of plasmatrough
  density derived from magnetospheric field line resonances, \textit{J.
  Geophys. Res.}, \textit{101}, 24,737--24,745, \doi{10.1029/96JA01083}, 1996.

\bibitem[{\textit{Wild et~al.}(2005)\textit{Wild, Yeoman, and Waters}}]{wild05}
Wild, J.~A., T.~K. Yeoman, and C.~L. Waters, Revised time of flight
  calculations for high latitude geomagnetic pulsations using a realistic
  magnetospheric magnetic field model, \textit{J. Geophys. Res.}, \textit{110},
  A11,206, \doi{10.1029/2004JA010964}, 2005.

\bibitem[{\textit{Zwan and Wolf}(1976)}]{zwan76}
Zwan, B.~J., and R.~A. Wolf, Depletion of solar wind plasma near a planetary
  boundary, \textit{J. Geophys. Res.}, \textit{81}, 1636--1648,
  \doi{10.1029/JA081i010p01636}, 1976.

\end{thebibliography}

\end{article}

\clearpage{}

\begin{table}
\begin{centering}
\begin{tabular}{|c|c|c|}
\hline 
Quantity & Model & Model Inputs\tabularnewline
\hline 
\multirow{1}{*}{$\mathbf{B}_{sph}$} & \multirow{1}{*}{T96} & $P_{dyn,sw}$, $B_{y,sw}$, $B_{z,sw}$, $Dst$, Dipole Tilt\tabularnewline
$\mathbf{B}_{msh}$ & KF94 & $\mathbf{B}_{sw}$, $r_{mp,0}$, $r_{bs,0}$\tabularnewline
\multirow{2}{*}{$n_{sph}$} & \multirow{2}{*}{Power Law $n=n_{0}\left(r_{mp,0}/r\right)^{m}$} & $m=2$ \citep{denton02}\tabularnewline
 &  & $n_{0}=$1~cm$^{-3}$ \citep[e.g.][]{lee96}\tabularnewline
$n_{msh}$ & \citet{spreiter66} & $n_{sw}$\tabularnewline
$u_{msh}$ & \citet{spreiter66} & $u_{sw}$\tabularnewline
$r_{mp,0}$ & \citet{shue98} & $P_{dyn,sw}$, $B_{z,sw}$\tabularnewline
$r_{bs,0}$ & \citet{landau59} & $r_{mp,0}$, $M_{ms}$\tabularnewline
\hline 
\end{tabular}
\par\end{centering}

\protect\caption{Summary of the models used in this study.\label{tab:Models}}
\end{table}

\begin{table}
\begin{centering}
\begin{tabular}{|c|c|c|c|}
\hline 
Variable & Med & IQR & $a_{i}$~(mHz)\tabularnewline
\hline 
$a_{0}$ & - & - & 0.687\tabularnewline
$B_{x,sw}$ & 0.16~nT & 5.03~nT & $\begin{cases}
-0.006 & B_{x,sw}<0\\
-0.006 & B_{x,sw}\geq0
\end{cases}$\tabularnewline
$B_{y,sw}$ & -0.16~nT & 5.00~nT & $\begin{cases}
+0.022 & B_{y,sw}<0\\
-0.027 & B_{y,sw}\geq0
\end{cases}$\tabularnewline
$B_{z,sw}$ & 0.03~nT & 3.47~nT & $\begin{cases}
-0.375 & B_{z,sw}<0\\
+0.031 & B_{z,sw}\geq0
\end{cases}$\tabularnewline
$Dst$ & -8~nT & 19~nT & $\begin{cases}
-0.199 & Dst<0\\
-0.112 & Dst\geq0
\end{cases}$\tabularnewline
$u_{sw}$ & 413~km~s$^{-1}$ & 142~km~s$^{-1}$ & 0.299\tabularnewline
$n_{sw}$ & 4.45~cm$^{-3}$ & 4.08~cm$^{-3}$ & 0.013\tabularnewline
$P_{dyn,sw}$ & 1.60~nPa & 1.26~nPa & 0.018\tabularnewline
\hline 
\end{tabular}
\par\end{centering}

\protect\caption{Results of the multiple linear regression model defined in Equation~\ref{eq:regression}.
The standard deviation of the residuals was 0.25~mHz.\label{tab:regression-results}}
\end{table}

\begin{table}
\begin{centering}
\begin{tabular}{|c|c|c|}
\hline 
Approximation & Region of inapplicability & Effect on $f_{KS}$\tabularnewline
\hline 
Model Quantities & - & $\pm10\%$\tabularnewline
WKB & - & $+15\pm4\%$\tabularnewline
$k_{\phi}=0$ & - & $-0.5\pm0.4\%$\tabularnewline
Incompressibility & Magnetosheath & $-20\pm4\%$\tabularnewline
No PDL & Low $\theta_{B}$ \& $M_{A}$ ($\sim1\%$ of time) & $+60\pm20\%$\tabularnewline
No Reconnection & $<64\%$ of time for $B_{z,sw}<0$ & Unsupported\tabularnewline
\hline 
\multicolumn{2}{|c|}{Overall Accuracy of Model Calculations} & $\begin{array}{c}
+4\\
-19
\end{array}\%$\tabularnewline
\hline 
\end{tabular}
\par\end{centering}

\protect\caption{Summary of the approximations in our model calculations and estimates
of their effects (including the spread) on the fundamental frequency
calculations.\label{tab:approximations}}
\end{table}

\end{document}